\documentstyle[epsf,aps,prl,twoside]{revtex} 
\input epsf.tex

\begin{document}
\draft
\twocolumn[\hsize\textwidth\columnwidth\hsize\csname@twocolumnfalse%
\endcsname

\title{\ \\Finite-Size Scaling Study of the Surface and Bulk 
 Critical Behavior\\  in the  Random-Bond $8$-state Potts Model}

\author{Christophe Chatelain 
and Bertrand Berche\cite{byline2}}

\address{Laboratoire de Physique des Mat\'eriaux,~\cite{byline3} 
Universit\'e Henri Poincar\'e, Nancy
1, B.P. 239,\\ 
F-54506  Vand\oe uvre les Nancy Cedex, France} 

\date{October 27, 1997, to appear in Phys. Rev. Lett.}

\maketitle

\begin{abstract}
The self-dual  
random-bond  eight-state Potts model is studied numerically through large-scale 
Monte Carlo 
simulations using the Swendsen-Wang cluster flipping algorithm. We
compute bulk and surface order parameters and susceptibilities and deduce 
the corresponding critical exponents at the random fixed point using standard
finite-size scaling techniques. The scaling 
laws are suitably satisfied. We find that a belonging of the model to the
2D Ising model universality class can be conclusively ruled out, and the 
dimensions of the relevant bulk and surface
scaling fields are found to
take the values $y_h=1.849$, $y_t=0.977$, $y_{h_s}=0.54$,  to be 
compared to their 
Ising values: $15/8$, $1$, and $1/2$.
\end{abstract} 

]

The understanding of the role played by impurities on the nature of phase transitions
is of great importance, both from experimental and theoretical perspectives.
It is a quite active field of research where the resort to large-scale
Monte Carlo simulations is often necessary~\cite{selkeshchurtalapov94}.
The effect of quenched bond randomness in a system which undergoes,  
in the homogeneous case, 
a second-order phase transition  has been extensively 
studied. It
is well understood since Harris proposed a relevance criterion for the
fluctuating interactions~\cite{harris74}.
Disorder appears to be a relevant perturbation  when the specific heat exponent
$\alpha$ of the pure system is positive. Since in the two-dimensional Ising
model (IM) $\alpha$ vanishes due to the logarithmic Onsager singularity, this
model was carefully studied in the '80s~\cite{shalaev94}. 

The analogous situation when the pure system exhibits a first-order transition
was less well studied, in spite of the early work of Imry and Wortis who
argued that quenched disorder could induce a second-order phase 
transition~\cite{imrywortis79}. This argument was then rigourously proved
by Aizenman and Wehr, and Hui and 
Berker~\cite{aizenmanwehr89huiberker89}. In two dimensions, even an 
infinitesimal amount of
quenched impurities changes the transition into a second-order one.
The first intensive Monte Carlo study of  the effect of disorder 
at a first-order 
phase transition is due to Chen, Ferrenberg and 
Landau (CFL).
These authors studied the $q=8$-state two-dimensional Potts model, which, 
 in the pure case,  is known to exhibit a first-order phase 
transition when $q>4$~\cite{wu82}.
They definitively showed that the transition  becomes
second-order in the presence of bond
randomness~\cite{chenferrenberglandau92}, and   obtained the critical
exponents from a finite-size scaling (FSS) study at the critical point of a 
self-dual disordered system~\cite{chenferrenberglandau95}.
Their results, together with other related 
works~\cite{novotnylandau81kardaretal95,wisemandomany95}, suggested that any 
two-dimensional
random system should belong to the
2D pure IM universality class~\cite{cardy96b}.
In a recent paper, Cardy and Jacobsen (CJ) used a different 
approach~\cite{cardyjacobsen97}, based on a transfer matrix 
formalism~\cite{blotenightingale82},
to study random-bond Potts models for different values of $q$. Their 
estimation
of the critical exponents  leads to a continuous variation of $\beta_b/\nu$ 
with $q$. This result is in accordance with previous theoretical calculations 
when $q\leq 4$~\cite{dotsenkopiccopujol95a},
and, in the randomness-induced second-order
phase transition regime $q>4$, it is quite different from the Ising
value of ${1\over 8}$, and in sharp disagreement with the Monte Carlo 
results of Ref.~\cite{chenferrenberglandau95} for $q=8$.

The surface properties of dilute magnetic systems paid less attention. 
Quite generally, the scaling laws involving  surface and/or bulk exponents 
can be deduced
from a homogeneity assumption for both surface and bulk singular free 
energies e.g.
\begin{equation}
f_{surf}(t,h,h_s)=b^{-(d-1)}f_{surf}(b^{y_t}t,b^{y_h}h,b^{y_{h_s}}h_s).
\label{eq-homogeneitys}\end{equation}
All the standard critical exponents can be expressed in terms of the 
anomalous dimensions $y_i$ associated to the relevant scaling fields~\cite{binder83}.
This makes their determination of great importance in the case of
random systems.
The
$(1,1)$ surface of the Ising model on a square lattice has only recently been 
investigated through Monte Carlo simulations by Selke et al~\cite{selkeetal97}.
The critical exponent $\beta_1$ of the surface magnetization was found to be 
very close 
to its value in the pure 2D IM.

In this letter, we report a FSS study of the bulk and surface 
critical properties
of the 8-state random-bond Potts model. 
 Although this model has already been studied by  Monte Carlo 
 simulations, our approach is the 
first investigation
of the surface properties for a random system other than Ising-like. It leads 
furthermore  to different results in which concerns the bulk properties, and 
our aim is to bring some clear evidence to solve the discrepancy between
the recent results of CJ (Ref.~\cite{cardyjacobsen97}) and those of CFL
(Ref.~\cite{chenferrenberglandau95}).

\begin{figure}
\epsfxsize=9cm
\begin{center}
\mbox{\epsfbox{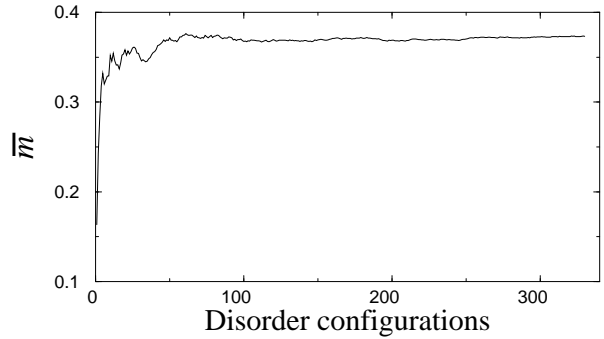}}
\end{center}\vskip -0.5cm
\caption{Fluctuations of the averages over the number of realizations of 
disorder. Average total magnetization for $L=56$ over up to 330 replicas.}\label{fluc}  
\end{figure}

In the following, we consider the
$q=8$-state random-bond Potts model on the square lattice. 
The Hamiltonian of the system with quenched random interactions is written
\begin{equation}
-\beta {\cal H}=\sum_{(i,j)}K_{ij}\delta_{\sigma_i,\sigma_j} 
\label{eq-hamPotts}
\end{equation}
where the spins  take the values $\sigma=1,2,\dots,q$ 
and the sum goes over nearest-neighbor pairs $(i,j)$.
The coupling 
strengths
are allowed to take two different values $K_1=K$ and $K_2=Kr$ with probabilities
$p$ and $1-p$ respectively. The ratio $K_2/K_1$ is kept
to the constant value $r=10$ (a strong enough value in order to ensure that the
critical behavior is no longer governed by the pure system fixed point) for 
all the simulations. If both couplings occur with the same 
probability, $p=0.5$, 
the system is, on average, self-dual, and the critical point is exactly given 
by the critical 
line
of the usual anisotropic model~\cite{wisemandomany95,fisch78kinzeldomany81}:
$({\rm e}^{K_c}-1)({\rm e}^{K_cr}-1)=q$. 
At $p=0.5$, we performed large-scale simulations of $L\times L$ lattices 
($10\leq L\leq 96$)
with periodic boundary conditions in one direction (vertical direction) and 
free boundaries (FBC system) in the 
other  direction. An equal number of couplings of both type is
first distributed over all the bonds of the lattice. The couplings are then 
mixed randomly. This procedure ensures an exact probability $p=0.5$, and avoids
the fluctuations around this value which would result from the
use of a random number generator to build the  
distributions of couplings.
The  multi-spin coding 
and the Swendsen-Wang 
cluster flipping method~\cite{swendsenwang87} 
were used, and the histogram technique allowed us to determine the behavior of
the different quantities over a range of 
$K$~\cite{ferrenbergswendsen8889}. For each distribution of
the couplings, between $2\times 10^5$ (smaller  lattice sizes)
 to $4\times 10^5$ (larger  lattice sizes) Monte Carlo steps per spin were 
 performed. Although it is smaller than the calculations of 
 Ref.~\cite{chenferrenberglandau95}, this is always larger than $10^4$ times 
 the correlation time, and turns out to be sufficient in order to produce reliable
 thermal averages.
 On the other hand, around 30 disorder realizations were performed in  
 Ref.~\cite{chenferrenberglandau95}, but, since 
 the  averages over randomness are still strongly fluctuating 
 (Fig.~\ref{fluc}), we used 
   500  
 ($10\leq L\leq 32$), 330 ($40\leq L\leq 64$), and
 250 ($72\leq L\leq 96$)
 disorder realizations. (For smaller lattice sizes, the configurational 
 fluctuations 
 of data due to randomness are more pronounced, and a larger number of
 configurations is needed). These values guarantee
 the same order of magnitude for the contributions to the statistical errors
 resulting from the thermal average and from the replica average.

The translational invariance is restored in the vertical direction 
by  averaging over the disorder realizations.
The local order parameter for a given replica, written $\mu_j$, 
 is  defined by the majority orientation of the spins 
at
column $j$~\cite{challalandaubinder86}:
\begin{equation}
\mu_j=\left(\frac{q\rho_{max}(j)-1}{q-1}\right).
\label{eq-locordpara}
\end{equation}
Here, $\rho_{max}(j)=max_\sigma(\rho_\sigma(j))$, where $\rho_\sigma(j)$ 
is the  density of spins in the state $\sigma$ at column $j$. The thermal 
average over the Monte Carlo iterations, written with brackets $\langle \dots
\rangle$, is performed and the physical 
quantities are then  averaged
over disorder configurations, for example 
$m_j\equiv\left[\langle \mu_j\rangle\right]$, where $\left[ \dots
\right]$ denotes the replica average.
The local surface
susceptibility  is given by $\chi_{11}=
KL\left[\langle \mu_1^{\ 2}\rangle-\langle \mu_1\rangle^2\right]$ and similar
quantities are defined for the bulk.

\vskip -2.8cm 

\begin{figure}
\epsfxsize=9cm
\begin{center}
\mbox{\epsfbox{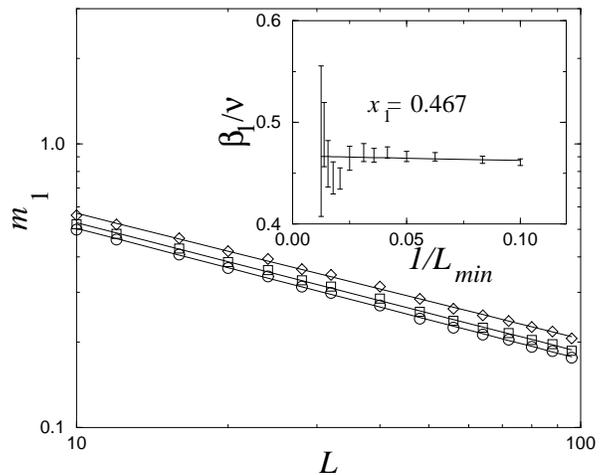}}
\end{center}\vskip -0.5cm
\caption{Log-log plot of $\left[\langle \mu_1\rangle\right]$ (circles), 
$\left[\langle \mu_1^{\ 2}\rangle\right]^{1/2}$ (squares), and 
$\left[\langle \mu_1^{\ 4}\rangle\right]^{1/4}$ (diamonds) vs $L$. The slopes over the whole
range of values of $L$ are respectively
$-0.461$, $-0.456$, and $-0.443$.
The insert shows the effective surface exponent $x_1(L_{min})$ defined in the text 
vs $L^{-1}_{min}$ 
(deduced from
$m_1$) and its extrapolated value.}\label{fig-m1}  
\end{figure}

The first part of our analysis was to study the local magnetic 
surface properties. The local surface magnetization 
$m_1=\left[\langle \mu_1\rangle\right]$ is 
expected to follow the 
usual finite-size scaling behavior at the infinite lattice critical point:  
$m_1(K_c,L)\sim L^{-x_1}$,
where the critical
dimension $x_1={\beta_1/\nu}$ is  deduced from the size dependence of 
$m_1$, 
$\left[\langle \mu_1^{\ 2}\rangle\right]$, and 
$\left[\langle \mu_1^{\ 4}\rangle\right]$ (Fig.~\ref{fig-m1}).

\vskip -0.5cm

A power-law fit of the curve between a given smaller size $L_{min}$ and 
the maximal 
value $L_{max}=96$ defines
an effective
exponent $x_1(L_{min})$. The smaller size is then cancelled from the data 
and the whole procedure is 
repeated until the three larger sizes only remain. The effective exponent is 
plotted against $L^{-1}_{min}$ (Insert in Fig.~\ref{fig-m1}), and 
the critical exponent follows
from the  extrapolation at infinite size in the linear regime. Here 
 the final estimation gives:
\begin{equation}
\frac{\beta_1}{\nu}=0.467\pm 0.006,\label{eq-beta1}
\end{equation}
where the uncertainty is the standard deviation.

\begin{figure}
\epsfxsize=9cm
\begin{center}
\mbox{\epsfbox{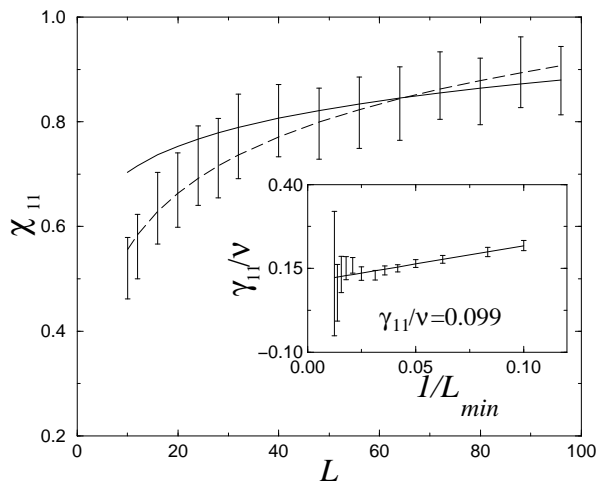}}
\end{center}\vskip -0.5cm
\caption{Local susceptibility $\chi_{11}(K_c)$ at the critical point vs $L$ and 
power-law fit 
(solid line) or logarithmic fit
(dashed line). The insert shows the effective exponent $\gamma_{11}/\nu$.}
\label{fig-chi11}  
\end{figure}

The behavior of the local surface susceptibility $\chi_{11}$ is more ambiguous,
since
$\chi_{11}$ seems to
 exhibit a   power-law
behavior with a very small exponent, but also fits with a
logarithmic divergence as it is the case for the pure IM. 
From the behaviors of
$\chi_{11}(K_c)$ and $\chi_{11}^{max}$ (deduced from histogram
reweighting), we obtain
$\gamma_{11}/\nu=0.099\pm 0.009$  (Fig.~\ref{fig-chi11}). It corresponds 
to a fit which gives greater 
place to large sizes.  On the other hand a logarithmic behavior seems also 
convincing (Fig.~\ref{fig-chi11}). Although  this first analysis does not allow 
any definitive conclusion, the scaling relation 
$2\beta_1/\nu +\gamma_{11}/\nu=d-1$ is best satisfied, within error bars, 
by the power-law case and is furthermore satisfying with the value of $x_1$
which rules out the 2D IM universality class.

\vskip -1.1cm

\begin{figure}
\epsfxsize=9cm
\begin{center}
\mbox{\epsfbox{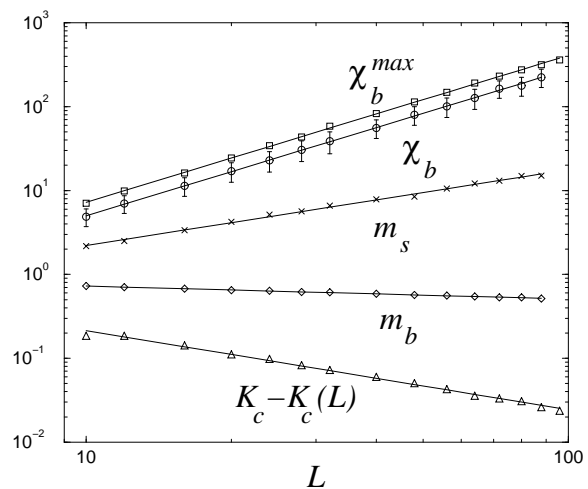}}
\end{center}\vskip -0.5cm
\caption{Bulk quantities: Maximum of the susceptibility ($\chi_b^{max}$, squares)
and its value at $K_c$ ($\chi_b$, circles); average magnetization ($m_b$, diamonds);
deviation of the effective critical coupling from its exact value (triangles), 
and surface excess magnetization ($m_s$, crosses). The 
corresponding exponents
$\gamma_b/\nu$, $-\beta_b/\nu$,  $-1/\nu$, and $-\beta_s/\nu$ are given
in the text.}
\label{fig-bulk}  
\end{figure}

Once the local surface properties have been studied and since they strongly
suggest that the model does not belong to the 2D IM universality class, 
a carefull analysis of bulk properties is needed to bring definitive conclusions.  
Bulk properties are furthermore needed
in order to compute surface excess magnetization.
For this purpose, we made  simulations on $L\times L$ lattices with periodic boundary
conditions in both directions (PBC system). The 
average quantities 
over the whole system 
lead to the critical exponents associated to
the bulk magnetization and bulk susceptibility  (Fig.~\ref{fig-bulk}). 
The determination of the slopes in the log-log plots, and of the corresponding 
standard  deviations (of the order of $10\%$),  indicates large
fluctuations.  We then turned back to the ``effective exponent'' technique 
presented above to have  accurate estimations.
One thus obtains
\begin{equation}
\frac{\beta_b}{\nu}=0.153\pm 0.003,\quad\frac{\gamma_b}{\nu}=1.701\pm 0.008.
\label{eq-betagammab}
\end{equation}
The first value is very different from the IM value and from the result 
of Ref.~\cite{chenferrenberglandau95} (0.126)  and closer  to the 
result  of Ref.~\cite{cardyjacobsen97} (0.142). 
The second value satisfies to better than $0.7\%$ the scaling law resulting 
from Rushbrooke  
and  hyperscaling relations:
$ {\gamma_b}/{\nu}=d-2{\beta_b}/{\nu}$.
The  correlation length exponent $\nu$
 is deduced  from the deviation of the effective critical 
 coupling (at the maximum of the 
bulk susceptibility $\chi_b^{max}$) from its exact value, 
$| K_c(L)-K_c | \sim L^{-1/\nu}$ (Fig.~\ref{fig-bulk}).
It leads to the correlation length exponent 
$\nu=1.023\pm 0.020$ (Fig.~\ref{fig-bulk})~\cite{otherestimation}.

\vskip -1cm

Excess surface magnetization can be calculated by a comparison between the 
FBC and PBC systems: 
\begin{equation}
m_s=\frac{1}{2}\sum_{j=1}^L(m_b-m_j)\simeq\frac{1}{2} L (m_b-\bar m)
\label{eq-mexcess}
\end{equation}
where $\bar m$ is the average magnetization for the FBC system and $m_b$ for
 the PBC one. In Eq.~\ref{eq-mexcess}, the approximation symbol renders the
 possible difference between the majority spin orientation on a layer $j$ and
 its value for the whole system. It produces a small difference between
 mean magnetization and mean profile for the FBC system
 but on average the two quantities should scale
 the same way.
The corresponding exponent  obtained from FSS takes the value:
\begin{equation}
\frac{\beta_s}{\nu}=-0.852\pm 0.007.
\label{eq-betagammas}
\end{equation}
It is in accordance with 
the expected value resulting
from the scaling law $\beta_s/\nu=\beta_b/\nu-1=-0.847$, albeit  
 $m_s$, given by a difference,  is subject to strong 
fluctuations.

In this letter, we reported the results of large-scale Monte Carlo 
simulations~\cite{CPU}
of the 
surface and bulk critical 
behaviors
of a randomness-induced second-order phase transition in the 8-state Potts
model. 
Concerning the bulk critical exponents, there is a clear discrepancy between
our results and those of Ref.~\cite{chenferrenberglandau95} which were very close
to the pure IM values. The main difference between our procedure and 
 these previous simulations 
is due to our use of a larger space of allowed coupling
strengths.
The self-dual condition $p=1/2$
was indeed imposed by these authors in both directions. A possible weakening of 
randomness could result of this choice.
We furthermore generated a  number of disorder realisations 10 times larger, which
makes our results reliable.  On the other hand, our value
for $\beta_b/\nu$ is  slightly above Cardy and Jacobsen's result, while we
 used  the same order parameter than CFL.  The possible explanation 
suggested in CJ (Ref.~\cite{cardyjacobsen97}) for the disagreement with CFL  
(non-standard order parameter) has thus to be dismissed.

We have also reported here the first extensive numerical study of surface 
critical 
behavior in a randomness-induced second-order phase transition. While excess
magnetization offers an alternate determination of the scaling dimension of
the {\it bulk} magnetic field, local surface properties lead to the scaling 
dimension of a {\it surface}  field which is also relevant.

We can summerize all the results in a table of the anomalous dimensions 
of the relevant
scaling fields (Table~\ref{tab1}). The independent determinations 
of these values, very close
together,  give reliability to the results. 
The final estimations are the following: $y_t=0.977$, $y_h=1.849$, 
$y_{h_s}=0.54$.

\vbox{
\narrowtext
\begin{table}
\caption{Scaling dimensions of the bulk and surface fields and 
of the temperature deduced from the values of the critical exponents.
\label{tab1}}
\begin{tabular}{lcccccc}
  & $\beta_1/\nu$ & $\gamma_{11}/\nu$ & $\nu$ & $\beta_b/\nu$ & $\gamma_b/\nu$ &
  $\beta_s/\nu$  \\
\tableline
$y_{h_s}$ & 0.533  & 0.549 &--&--&--&--\\
$y_t$ &--&--&0.977&--&--&-- \\
$y_h$ &--&--&--&1.847&1.850&1.852 \\
\end{tabular}
\end{table}
\narrowtext
 
 }\vskip -5mm

Finally, one has to mention that we
 also computed profiles and correlations (details will be published 
elsewhere). The values of 
$\eta=0.29$ (correlations PBC system),  and of the critical exponents difference 
$x_1-x_b=0.27$ (profile close to the free surfaces, FBC system) lead to
results which are slightly too small compared to 
 the previous values of $x_1$ and $x_b$. Surprinsingly, the bulk
 exponent $x_b=0.145$ is found to be very close to Cardy and Jacobsen's result
 which was deduced from the behavior of correlations as well, but 
 within a strip geometry.

We thank the Ciril and the Centre 
 Charles Hermite in Nancy for computational facilities.


\begin{references}
\bibitem{otherestimation} Other estimations of $\nu$ can in principle be obtained by the size behavior of 
$\left.\frac{d\ln X}{d K}\right|_{K_c}$ where $X$ is some quantity 
(magnetization for 
example), but they lead to estimations of $\nu$ slightly below 1, while the
stability of the random fixed point requires $\nu\geq 1$.
\end{references}
\end{document}